\documentclass{vietnam}

\usepackage{amssymb}

\def\be{\begin{equation}}
\def\ee{\end{equation}}
\def\bea{\begin{eqnarray}}
\def\eea{\end{eqnarray}}

\newcommand{\Photo}{}

\begin{document}
\vspace*{4cm}
\title{Magnetic monopoles -- theory overview}

\author{ Arttu Rajantie }

\address{Department of Physics, Imperial College London,\\
London SW7 2AZ, United Kingdom}

\maketitle\abstracts{
I give a theoretical overview of magnetic monopoles, focusing on the physical perspective of monopoles as hypothetical particles rather than as mathematical objects. I argue that monopoles are exceptionally interesting hypothetical particles and discuss the prospects of addressing the question of their existence, and possibly producing them, in particle physics experiments.
}

\section{Introduction}
We learn early on in school that magnets always have two poles, north and south. In other words, magnetic monopoles, isolated magnetic north or south poles do not exist. This is in stark contrast with electricity, as particles with a positive or negative electric charge do exist. 

Classical electrodynamics provides no explanation for the apparent non-existence of magnetic monopoles. In his 1873 treatise,\cite{maxwell1873treatise} Maxwell makes the explicit assumtion that magnetic monopoles do not exist, or in his words, \emph{``In every Magnet the total quantity of Magnetism (reckoned algebraically) is zero.''} He justifies this statement using two pieces of empirical evidence:
\emph{``This may be easily proved by putting the magnet into a small vessel and floating it in water. 
	The vessel will turn in a certain direction, 
	...  but there will be no motion of the vessel as a whole in any direction; so that there can be no excess of the force towards the north over that towards the south, or the reverse. It may also be shewn from the fact that magnetizing a piece of steel does not alter its weight.''}

We have made a huge amount of progress in both theoretical and experimental physics in the 151 years since Maxwell wrote these words. It is therefore worth asking whether, and to what extent we really know that magnetic monopoles do not exist. 

In their conventional form, Maxwell's equations of classical electrodynamics describe the electric and magnetic fields in the presence of electric charges $\rho_E$ and currents $\vec{\jmath}_E$.
It is, however, entirely straightforward to generalise the equations to allow for magnetic charges $\rho_M$ and currents $\vec{\jmath}_M$,
\begin{eqnarray}
	\vec{\nabla}\cdot\vec{E} &=& \rho_E,\quad\quad 
		\vec{\nabla}\times\vec{E} = -\frac{\partial\vec{B}}{\partial t}-\vec{j}_M,
	\nonumber\\
	\vec{\nabla}\cdot\vec{B} &=& \rho_M,\quad \quad
	\vec{\nabla}\times\vec{B} = \frac{\partial\vec{E}}{\partial t}+\vec{\jmath}_E.
\end{eqnarray}
It has been often noted that as a consequence of this addition, the equations would acquire an additional electromagnetic duality symmetry
\begin{eqnarray}
	\label{equ:EMduality}
	\vec{E}+i\vec{B}&\rightarrow& e^{i\theta}\left(\vec{E}+i\vec{B}\right);
	\quad
	\rho_E+i\rho_M\rightarrow e^{i\theta}\left(\rho_E+i\rho_M\right);
	\quad
	\vec{\jmath}_E+i\vec{\jmath}_M\rightarrow e^{i\theta}\left(\vec{\jmath}_E+i\vec{\jmath}_M\right).
\end{eqnarray}
This symmetry implies that, at least in classical electrodynamics, the laws of electricity and magnetism are identical, and the only difference between them is the empirical statement that magnetic monopoles do not appear to exist but electric charges do.

\section{Dirac Monopole}

At first sight, quantum mechanics appears to provide an explanation for the absence of magnetic monopoles. While in classical physics it is possible to express everything in terms of $\vec{E}$ and $\vec{B}$, quantum mechanics requires the use of the scalar potential $\phi$ and the vector potential $\vec{A}$, so that $\vec{B}=\vec{\nabla}\times\vec{A}$.
Because $\vec{\nabla}\cdot\vec{\nabla}\times\vec{A}=0$ for any non-singular $\vec{A}$, it seems that this prohibits the existence of magnetic monopoles.

However, Paul Dirac showed in 1931 that this conclusion is premature.\cite{Dirac:1931kp} One can consider the singular 
\emph{Dirac monopole} configuration
\begin{equation}
	\label{equ:Diracmonopole}
	\vec{A}(\vec{r})=\frac{g}{4\pi r}\frac{\vec{r}\times\hat{n}}{r-\vec{r}\cdot\hat{n}},
\end{equation}
where $g$ is a real parameter and $\hat{n}$ is a constant unit vector. This configuration has a line singularity, known as the Dirac string, which extends from the origin to infinity in the direction of $\hat{n}$. Away from this singularity, the magnetic field it corresponds to is identical to that of a pointlike magnetic monopole with magnetic charge $g$ placed at the origin. Furthermore, Dirac showed that if all electric charges are integer multiples of $2\pi/g$, then the Dirac string is unobservable and can actually be moved by a (singular) gauge transformation. Therefore, Eq.~(\ref{equ:Diracmonopole}) physically describes a magnetic monopole.

This finding implies that quantum mechanics is compatible with the existence of magnetic monopoles if electric charge is quantised. As Dirac wrote in 1948, 
\emph{''The quantization of electricity is one of the most fundamental and striking features of atomic physics, and there seems to be no explanation for it apart from the theory of poles. This provides some grounds for believing in the existence of these poles.''~\cite{Dirac:1948um}}

The quantisation condition also implies that if magnetic monopoles exist, their magnetic charge $g$ must be an integer multiple of the Dirac charge
\begin{equation}
	\label{equ:Diraccharge}
	g_D=\frac{2\pi}{e_0}\approx 20.7,
\end{equation}
where $e_0$ is the elementary electric charge which I have assumed to be the charge of the proton. This is useful for experiments, because we know the strength of the magnetic charge we need to be looking for, and because $g_D\gg 1$ monopoles should be easy to detect if they exist.

It is possible to embed the Dirac monopole~(\ref{equ:Diracmonopole}) in the Standard Model. Because quarks have electric charges $-1/3$ and $2/3$, it may appear that the smallest magnetic charge allowed by the quantisation condition should be $3g_D$. However, it turns out that a single Dirac charge is possible if the monopole also carries ${\rm SU}(3)$ chromomagnetic charge.\cite{Corrigan:1976wk} Therefore the relevant charge $e_0$ in Eq.~(\ref{equ:Diraccharge}) is, indeed, the charge of the proton.

The Dirac monopole configuration (\ref{equ:Diracmonopole}) and its generalisations describe the electromagnetic field around a static magnetic monopole, in the same way as the Coulomb field describes the electric field around a static electric charge. It is a solution of the classical equations of motion  away from the origin, but not at the origin itself. Therefore it needs a source particle, just like the Coulomb field needs an electrically charged particle such as an electron as its source. In both electric and magnetic cases, the energy of the classical field configuration is divergent, but our experience with electrically charged particles shows that it does not mean that the physical mass of the particle is infinite. Nevertheless, one can modify the Standard Model Lagrangian in such a way that the Dirac field configuration (\ref{equ:Diracmonopole}) has a finite energy.\cite{Cho:1996qd,Cho:2012bq,Ellis:2016glu,Blaschke:2017pym} Interpreting this energy as the mass of the particle, it has been found that these \emph{Cho-Maison monopoles} could be as light as $2.37~{\rm TeV}$.\cite{Blaschke:2017pym}

A very interesting result, known as the \emph{Callan-Rubakov effect},\cite{Callan:1982ah,Callan:1982ac,Rubakov:1982fp} shows that when a light electrically charged fermion scatters off a magnetic monopole, its wave function can reach the monopole core with the effect that its helicity flips. Most strikingly, in the context of the Standard Model, this can lead to baryon number violating processes which would catalyse proton decay. Most importantly this process is not suppressed by the mass scale of the monopole, and therefore it would not require high collision energies. The Callan-Rubakov effect is still not fully understood because of the lack of a complete quantum field theory description and it is therefore an active area of research.\cite{Brennan:2023tae,vanBeest:2023mbs,Khoze:2024hlb}

Overall, magnetic monopoles are exceptionally interesting compared to other hypothetical particles that physicists are looking for. The Standard Model already describes how they would behave if they exist. As Dirac wrote in 1931, \emph{''This new development requires no change whatever inthe formalism when expressed in terms of abstract symbols denoting statesand observables, but is merely a generalisation of the possibilities of representation of these abstract symbols by wave functions and matrices. Under these circumstances one would be surprised if Nature had made no use of it.''}~\cite{Dirac:1931kp} The laws of electrodynamics imply that the lightest magnetically charged particle has to be stable, and because by definition it will interact with the electromagnetic field with a strong coupling constant, it should be possible to trap a magnetic monopole and use it for further experiments. Because of the Callan-Rubakov effect, those experiments would not require high energy collisions and would still provide a wealth of new information about fundamental physics.

\section{Monopoles in Quantum Field Theory}
\label{sec:QFT}

In order to describe the monopole itself, as opposed to other particles in its presence, one needs a quantum field theory of magnetic monopoles. Through the electromagnetic duality, one would expect that a theory with only monopoles and no electrically charged particles is equivalent to quantum electrodynamics with coupling constant $g_D$. However, for a complete understanding of the physics of magnetic monopoles one needs a quantum theory which has both electrically and magnetically charged particles.

In 1974 't~Hooft~\cite{tHooft:1974kcl} and Polyakov~\cite{Polyakov:1974ek} independently found that certain weakly coupled quantum field theories have magnetic monopoles in their particle spectrum, arising from semiclassical non-linear solutions of the field equations. The simplest example is the ${\rm SU}(2)$ gauge field theory with a scalar field $\phi$ in the adjoint representation. When the scalar field has a non-zero vacuum expectation value, it breaks the gauge symmetry spontaneously to ${\rm U}(1)$. Therefore, the low-energy effective theory is electrodynamics with a massless photon, a neutral Higgs boson, and electrically charged vector bosons, as well as the \emph{'t~Hooft-Polyakov monopole} which is described by the ``hedgehog'' configuration
$\phi^a(\vec{x}) = f(|\vec{x}|)x_a$. This is a quasiparticle with a finite size $R\sim m_v^{-1}$, where $m_v$ is the mass of the vector bosons, and a magnetic charge $g=4\pi/e$, where $e$ is the electric charge of the vector bosons. Its mass is given by the energy of the configuration as is approximately $M\sim m_v/e^2$. In a weakly coupled theory, with $e\ll 1$, it is therefore always heavier than the vector bosons.

In particular, simple topological arguments show that 't~Hooft-Polyakov monopoles exist in any Grand Unified Theories (GUTs) which unify the electroweak and strong force to a single gauge group. In typical GUTs, the unification scale is around $10^{16}~{\rm GeV}$, which means that the monopole mass would be $10^{17}~{\rm GeV}$. This is a very high mass, well beyond the reach of any foreseeable particle colliders. Analogous monopole solutions also exist in string theory, with even higher masses. If one believes in these theories one will therefore conclude that magnetic monopole exist in principle, as particle states in the spectrum of the ``theory of everything'', but they are so heavy that we cannot produce them.

There are some theories beyond the Standard Model that predict lighter monopoles, for example Pati-Salam theory~\cite{Pati:1974yy} or the trinification model,\cite{Glashow:1984gc} whose one specific variant~\cite{Raut:2022ryj} was recently found to predict monopoles with masses as low as $160~{\rm TeV}$.

Another approach to achieve electroweak-scale monopoles is by extending the matter sector of the Standard Model. The theory proposed by P.Q.\ Hung~\cite{Hung:2020vuo,Ellis:2020bpy} contains four scalar doublets, one real scalar triplet, and one complex scalar triplet. The non-zero vacuum expectation value of the scalar triplet gives rise to 't~Hooft-Polyakov -like \emph{Hung monopoles} with electroweak scale masses which could be as low as $900~{\rm GeV}$. 

Of course, magnetic monopoles do not have to be semiclassical quasiparticles. Instead, one can consider monopoles that are elementary particles, in the same way as the Standard Model particle but with a non-zero magnetic charge.
There are formulations of such a theory for elementary magnetic monopoles,\cite{Schwinger:1966nj,Zwanziger:1970hk}
but they are cumbersome to use for practical calculations. They also suffer from the problem that one cannot expect perturbation theory to be valid because the coupling constant $g_D$ is large. Nevertheless, these are likely to be merely practical inconveniences rather than genuine obstacles -- after all, the examples of semiclassical monopoles in well-defined and well-understood quantum field theories demonstrate that there cannot be any fundamental reason why quantum field theories could not allow monopole particles.

On the other hand, the difference between semiclassical and elementary monopoles may be illusory. It has been argued that because of quantum effects, even an elementary monopole, which appears in the theory as a pointlike particle, must have an effective size of $R\sim g^2/8\pi M$, where $g$ and $M$ are its magnetic charge and mass.\cite{goebel1970spatial,goldhaber1983monopoles} This is comparable to the size of the 't~Hooft-Polyakov monopole, so elementary monopoles may not be physically any more pointlike than semiclassical monopoles. In fact, many theories have a Montonen-Olive duality symmetry between electrically charged elementary particles and semiclassical monopole particles, which means that there cannot be any physical difference between them.~\cite{Montonen:1977sn}

If magnetic monopoles are elementary particles, then their mass is a free parameter which needs to be determined by experiment, just like masses of other elementary particles, and not calculable from the theory. This means that it does not have to be high, and one can meaningfully ask whether monopoles with masses within the reach of our current particle physics experiments exist.

\section{Monopole Searches}

There are two types of empirical searches for magnetic monopoles. Of the first type are the searches attempting to find monopoles that are physically present in the Universe today. This includes those described by Maxwell in 1873 as well as many others in the last one and a half century. Monopoles have been searched in the bottom of the oceans,\cite{Fleischer:1969qyz,Kolm:1971xb}, moon rocks,\cite{Alvarez:1970zu} iron ores,\cite{Ebisu:1986dw}, polar volcanic rocks,\cite{Bendtz:2013tj} and many other samples of matter.\cite{Jeon:1995rf} Cosmic ray experiments, most notably MACRO,\cite{MACRO:2002jdv} but also others such as IceCube~\cite{IceCube:2021eye} and ANTARES~\cite{ANTARES:2022zbr} recently, have carried out monopole searches. Because none of these searches have found monopoles, they place an upper bound on the flux of magnetic monopoles of approximately $10^{-16}\ {\rm cm^{-2}s^{-1}sr^{-1}}$ in general and $10^{-19}\ {\rm cm^{-2}s^{-1}sr^{-1}}$ for highly relativistic monopoles.
The abundance of magnetic monopoles in galaxies is also constrained by the existence of the galactic magnetic fields.\cite{Parker:1970xv,Adams:1993fj}

The second type of searches are testing the hypothesis that even if there are no or very few monopoles present in the Universe today, they may still exist as particle states that could be produced in experiments. Particle physicists have been looking for monopoles in  all major particle physics experiments, including the Large Hadron Collider {LHC} at CERN, where these searches have been carried out by the MoEDAL and ATLAS experiments. Because of their very different designs, ATLAS tends to be more sensitive to monopoles with one or two Dirac charges, and MoEDAL to monopoles with a higher charge. No monopoles have been found so far, which places upper bounds on the monopole production cross section. If one has a theoretical prediction of the cross section as a function of the monopole mass, these then imply lower bounds on the monopole mass.

In proton-proton collisions, monopole production has been assumed to take place through the tree-level Drell-Yan or photon fusion processes. The mass limits inferred from these experiments are approximately $3$ to $3.5~{\rm TeV}$ depending on the spin and the magnetic charge of the particle.\cite{ATLAS:2023esy,MoEDAL:2023ost} However, one should not take these bounds literally because of the strong coupling constant $g_D\approx 20.7\gg 1$, which appears as the expansion parameter in perturbation theory. It means that the loop corrections are likely to be larger than the tree-level term and increase order by order, so that the loop expansion does not converge.

In fact, the problem could be much more serious than this, at least for 't~Hooft-Polyakov monopoles. There are solid theoretical arguments that the production of semiclassical states, such as 't~Hooft-Polyakov monopoles, is suppressed by the exponentially small factor $\exp(-4/\alpha)\sim 10^{-238}$, where $\alpha$ is the fine structure constant.\cite{Witten:1979kh,Drukier:1981fq} This can be understood as a consequence of entropy: A high-energy collision produces a large number of quanta, and in order to produce a semiclassical monopole they would have to be arranged in a very specific, ordered way, and the probability of this is $O(e^{-4/\alpha})$.
This exponential suppression has been demonstrated by a numerical calculation for kink pair production in 1+1 dimensions,~\cite{Demidov:2011dk} but not for magnetic monopoles yet. Nevertheless, if true, it would mean that it is practically impossible to produce semiclassical monopoles in proton-proton collisions and therefore these experiments cannot provide any bounds on the monopole mass.

In itself, the argument about the exponential suppression applies only to semiclassical monopoles, and therefore one can hope that the production cross section of elementary monopoles does not suffer from the same suppression. However, as discussed in Section~\ref{sec:QFT}, there are good reasons to believe that there may not be an actual physical difference between the properties of elementary and semiclassical monopoles, so that hope may not be justified.

\section{Dual Schwinger Process}

If production of monopoles in collisions of elementary particles is not possible, then one needs to look at alternative setups that would not be subject to the same suppression and where the production process can be described without the use of perturbation theory. One candidate is the dual Schwinger process.\cite{Affleck:1981ag,Gould:2017zwi}
 
The Schwinger process itself refers to the production of electrically charged particle-anti\-par\-ticle pairs in a strong electric field.\cite{Sauter:1931zz,Schwinger:1951nm} It can be understood as quantum tunneling through the Coulomb barrier and its rate can be computed using instanton techniques which do not require a weak coupling.\cite{Affleck:1981ag,Affleck:1981bma} Through the electromagnetic duality (\ref{equ:EMduality}), the same result gives the pair production of rate of magnetic monopoles with mass $M$ and magnetic charge $g$ per unit time and volume in a constant and uniform magnetic field $\vec{B}$,
\begin{equation}
	\Gamma=\frac{g^2|\vec{B}|^2}{8\pi^3}e^{-\frac{\pi M^2}{g|\vec{B}|}+\frac{g^2}{4}}.
\end{equation}
This implies that pair production will be unsuppressed if $M\lesssim (g^3|\vec{B}|/4\pi)^{1/2}$. For example, the LHC magnets have a field strength of $|\vec{B}|\sim 10~{\rm T}$, so if magnetic monopoles with mass less than a few ${\rm keV}$ existed, they would have been produced in large numbers as soon as LHC was switched on.

Stronger mass bounds are obtained by considering magnetars, neutron stars with exceptionally strong magnetic fields of up to $10^{11}~{\rm T}$. If magnetic monopoles with masses less than roughly $1~{\rm GeV}$ existed, they would be produced by the dual Schwinger process outside the surface of the star. The magnetic field would accelerate them along the field lines, reducing the strength of the field. The fact that the fields still survive implies that this has not happened, and therefore places a lower bound on the monopole mass.\cite{Gould:2017zwi} In the interior of the star, the field can be stronger, leading to stronger bounds. On the other hand, it has also recently been suggested that if monopoles with a suitable mass exist, their production by the dual Schwinger process could explain the observed magnetic field fluctuations of magnetars.\cite{Klyuev:2024vzm}

The strongest known magnetic fields in the Universe, and therefore the best prospects for production of magnetic monopoles through the dual Schwinger process, exist in heavy ion collisions at the LHC. In ultraperipheral collisions, the magnetic field becomes as strong as $B\sim 10^{16}~{\rm T}$ is the space between the ions, although only for a short time of $\omega^{-1}$ with $\omega\approx 73~{\rm GeV}$. Because of the strong time dependence, calculating the production probability is significantly harder than for a constant field, but worldline instanton calculations provide a conservative estimate for the production cross section~\cite{Gould:2019myj,Gould:2021bre,MoEDAL:2021vix}
\begin{equation}
	\label{equ:sigmaestimate}
	\sigma\gtrsim \frac{2(gB)^4 R^4}{9\pi^2 M^5\omega}e^{-4M/\omega},
\end{equation}
where $R$ is the radius of the colliding nuclei, and $g$ and $M$ are the magnetic charge and mass of the monopole. Based on this, MoEDAL has carried out monopole searches in heavy ion collisions in LHC Run 2  using aluminium trapping detectors,\cite{MoEDAL:2021vix} and  later also using the CMS beam pipe which was exposed to heavy ion collisions in Run 1.\cite{MoEDAL:2024wbc} Together, these place a lower bound of approximately $80~{\rm GeV}$ on the monopole mass. 
ATLAS has published a similar search based on Run 3, exclusing masses of up to $120~{\rm GeV}$ for monopoles with a single Dirac charge.\cite{ATLAS:2024nzp} 

The relevant parameters characterising the collision in Eq.~(\ref{equ:sigmaestimate}), $B$ and $\omega$, both scale linearly with the collision energy, and therefore mass reach of heavy ion collision experiments will also scale nearly linearly with the collision energy. Using the Future Circular Collider with collision energy per nucleon of $100~{\rm TeV}$ one can therefore expect to reach monopole masses $M\sim 1.5~{\rm TeV}$, which may be enough to produce Hung monopoles if they exist. Looking even further into the future, it has been argued that the most powerful collider that could be built using extensions of existing technology would be a 11000 km circular collider on the Moon,\cite{Beacham:2021lgt}
with a collision energy of around $10~{\rm PeV}$ per nucleon in heavy ion collisions. That could be enough for producing magnetic monopoles of masses around $150~{\rm TeV}$ which are predicted by trinification models.

A great deal more theoretical work is also needed to make sure that the predictions are applicable to such extreme collision energies. On the other hand, Eq.~(\ref{equ:sigmaestimate}) is a conservative lower bound, and improved theoretical calculations are therefore expected to strengthen the mass bounds. Some work has already been done: For example, full field theory instanton calculations show that at least in a constant field, the production rate of 't~Hooft-Polyakov monopoles is higher than the worldline instanton calculation suggests.\cite{Ho:2019ads,Ho:2021uem} 

\section*{Acknowledgments}

I would like to thank the organisers of PASCOS-24 for a fantastic conference. In particular, I would like to dedicate this paper to the memory of P.Q. Hung, the chair of the conference and a brilliant physicist who sadly passed away on 
7 October 2024.

\end{document}